\documentclass[sigconf,nonacm]{acmart}
\usepackage[T1]{fontenc}
\usepackage{tabularx,booktabs}
\usepackage{makecell}
\usepackage{tcolorbox}
\usepackage{url}
\usepackage{soul}
\usepackage{enumitem}

\newcommand{\update}[1]{\textcolor{black}{#1}}

\acmConference[EASE 2025]{The 29th International Conference on Evaluation and Assessment in Software Engineering}{17–20 June, 2025}{Istanbul, Türkiye}

\begin{document}

\title{\update{Towards User-Centred Design of AI-Assisted Decision-Making in Law Enforcement}}

%



\author{Vesna Nowack}
\author{Dalal Alrajeh}
\affiliation{  
  \institution{Imperial College London, UK}
  \city{}
  \country{}
}
\email{{v.nowack,dalal.alrajeh}@imperial.ac.uk}

\author{\hspace{1mm}Carolina Gutierrez Mu\~{n}oz}
\author{\hspace{1mm}Katie Thomas}
\author{\hspace{1mm}William Hobson}
\affiliation{  \institution{\hspace{1mm}University of Bath, UK}
\city{}
\country{}}
   \email{{cigm20,kt697}@bath.ac.uk}
   \email{wghobson1@gmail.com}

\author{Patrick Benjamin\footnotemark[1]}
\affiliation{  
  \institution{University of Oxford, UK}
  \city{}
  \country{}
}
\email{patrick.benjamin@jesus.ox.ac.uk}

\author{Catherine Hamilton-Giachritsis}
\affiliation{  \institution{University of Bath, UK}
\city{}
\country{}}
   \email{chg26@bath.ac.uk}

\author{Tim Grant}
\affiliation{  \institution{Aston University, UK}
\city{}
\country{}}
    \email{t.d.grant@aston.ac.uk}

\author{Juliane A. Kloess\footnotemark[2]}
\affiliation{  \institution{University of Edinburgh, UK}
\city{}
\country{}}
    \email{juliane.kloess@ed.ac.uk}

\author{Jessica Woodhams}
\affiliation{  \institution{University of Birmingham, UK}
\city{}
\country{}}
    \email{j.woodhams@bham.ac.uk}

\renewcommand{\shortauthors}{Nowack et al.}

\begin{abstract}
Artificial Intelligence (AI) has become an important part of our everyday lives, yet user requirements for designing 
AI-assisted systems in law enforcement remain unclear.
To address this gap, we conducted qualitative research on decision-making
within a law enforcement agency. Our study aimed to identify limitations of existing practices, explore user 
requirements and understand the responsibilities that humans expect to undertake in these systems.

Participants in our study highlighted the need for a system capable of processing and analysing large volumes of data efficiently to help
in crime detection and prevention. Additionally, the system should satisfy requirements for 
scalability, accuracy, justification, trustworthiness and adaptability to be adopted in this domain.
Participants also emphasised the importance of having end users review the input data that might be challenging for AI to interpret,
and validate the generated output to ensure the system's accuracy. To keep up with the evolving nature of the law enforcement domain, end users need to help the system adapt to the changes 
in criminal behaviour and government guidance, and technical experts need to regularly oversee and monitor the system. Furthermore, user-friendly human interaction with the system is essential for its adoption 
and some of the participants confirmed they would be happy to be in the loop and provide necessary feedback that the system can learn from. 
Finally, we argue that it is very unlikely that the system will ever achieve full automation due to the dynamic and complex nature of the law enforcement domain.

\end{abstract}


\maketitle

\footnotetext[1]{Patrick Benjamin was at Imperial College London when this work was undertaken.}
\footnotetext[2]{Juliane A. Kloess was at University of Birmingham when this work was undertaken.}

\vspace{-2mm}\section{Introduction}
\label{section:introduction}



Artificial intelligence (AI) is becoming increasingly influential across various sectors of society, including the policing and security services. 
The potential benefits of AI in policing are 
numerous, ranging from preventing and solving crimes~\cite{mcdaniel2021predictive,montasari2023countering}, to addressing the evolving nature of criminal 
activities~\cite{UKgovLetter2023}. Several police forces have employed `predictive policing' tools, which utilise algorithms and historical data to forecast the probable locations of specific crimes, such as burglaries and street violence~\cite{Christie2021}.
Limited use has also included assessing the likelihood of individuals exhibiting certain behaviours or traits in the future~\cite{custers2022ai}.
Nonetheless, the adoption of AI in policing faces challenges. A primary concern is the potential for algorithms to introduce, replicate, or amplify biases~\cite{SANTOW2024}.

One response to such concerns in the United Kingdom (UK) has been the development of the 'Covenant for Using Artificial Intelligence (AI) in Policing' by the UK National Police Chiefs' Council~\cite{NPCC2023}. 
The Covenant establishes a set of principles that police forces have committed to follow in their employment of AI technologies. As a result of this endorsement, AI developers and users within 
the policing sector are required to adhere to the Covenant's guidelines around a lawful, transparent, explainable, responsible, accountable and robust AI design. In a similar vain,  both the European Commission 
and the EU High-Level Expert Group on AI emphasised that the division of responsibilities between humans and AI systems should adhere to human-centric design principles, ensuring meaningful 
opportunities for human intervention and decision-making~\cite{Enarsson2021ApproachingTH}. 


This means that the design of AI-based systems in law enforcement should allow for humans to maintain oversight and interact where needed. 
A user-centred design approach would need to address humans' concerns \update{regarding the use of} AI early in the design process. 
Gaining insight into human perspectives and capabilities is essential
for creating systems that are both effective and trusted. \update{By understanding user needs from the start}, potential challenges such as biases (AI and human), misuse, or lack of transparency,
\update{as well as unfairness and privacy issues~\cite{Kong2024Toward},} 
can be mitigated, 
ensuring that AI tools align with end users' expectations and the ethical principles \update{outlined in the Covenant}. Additionally, \update{it is important to understand what responsibilities humans may undertake} 
to guarantee that the decisions in AI-assisted decision-making
are made according to the law and official guidance~\cite{Enarsson2021ApproachingTH}.



While some \update{user} studies demonstrate practitioners' attitudes and an early experience towards AI-based systems in technology companies
\cite{Winter2022Towards,Williams2024UserCentric,Kang2024AQuantitative}, little is known about users' opinions of
AI assisting them in everyday activities in law enforcement agencies (LEAs). A study by Dempsey et al.~\cite{Dempsey2023Exploring} indicates mixed views
of police officers towards AI and the authors argue that the features, such as fairness, accountability, transparency and explainability,
are needed for a ``responsible design of AI policing technologies''.
On the other hand, Herrewijnen et al.~\cite{Herrewijnen2024Requirements} present
positive attitudes of police officers towards AI. Their study highlights the importance of the human involvement, 
but does not \update{identify or explore the specific responsibilities that humans need to undertake in this process}.


Consequently, we conducted a \update{user} study with criminal investigators (\textit{investigators} for short) from a LEA in the UK to \update{learn about their current 
decision-making practices}, understand their perspectives on using an AI-assisted system to support their tasks and to define the roles they need to take \update{within} such a system. 
The aim of this study was to \update{address the following research questions:}
\vspace{2mm}

\noindent \textbf{RQ1:} \update{What are the key user needs driving the design of an AI-assisted decision-making system in law enforcement?}\\
\noindent \textbf{RQ2:} What are the \update{primary} quality requirements for an AI-assisted decision-making system in law enforcement?\\
\noindent \textbf{RQ3:} \update{What are the responsibilities humans need to undertake in an AI-assisted decision-making system in law enforcement?}\\


\vspace{-2mm}

\noindent The contributions of this paper are as follows:
\vspace{-2mm}

\begin{itemize}[leftmargin=10pt]
    \item To the best of our knowledge, this is the first in-depth require\-ments-based qualitative analysis of investigators' perspectives 
    on an AI-assisted decision-making system in a LEA. Based on the requirements elicited from interviews, 
    we offer insights into the limitations of the current decision-making tools in use and highlight a need for 
    a tool that would overcome these limitations and help users with their everyday work.
    \item Our findings identify the essential quality requirements that an AI-assisted system needs to satisfy to be successfully adopted and 
    accepted in a LEA. Scalability, accuracy, justification, trustworthiness, adaptability and user-friendly interaction
    were recognised as the most important requirements for the system's adoption.
    \item Our findings highlight the critical importance of having humans in the loop, discussing the various \update{responsibilities} and the necessary expertise
    needed to \update{validate} input and output data, and to maintain an AI-assisted system during the deployment. We also argue that it is 
    very unlikely that the system will ever achieve full automation due to the dynamic and complex nature of the law enforcement domain.
\end{itemize}

The rest of the paper is structured as follows. The next section provides the background of the collaboration between the academic institutions and the LEA. 
We describe the methodology of our qualitative research in Section~\ref{section:methodology} and our findings in
Section~\ref{section:findings}. The design of our system is presented in Section~\ref{section:prototype}. 
Further discussions can be found in Section~\ref{section:discussion} and threats to validity in Section~\ref{section:threats}.
We summarise the related work in Section~\ref{section:related} and conclude in Section~\ref{section:conclusion}.
\section{Background}
\label{section:background}


This study was conducted as part of a five-year collaboration between a UK LEA and four UK Universities (Imperial College London, Bath, Birmingham and Aston). 
The general aim was to investigate AI-based methods and address socio-technical challenges associated with designing AI-assisted software 
that would help investigators reduce their manual work when analysing potentially distressing criminal material and making decisions around detected crimes.


This multidisciplinary team brought together experts from software engineering, forensic and clinical psychology, and forensic linguistics. At the beginning, 
the team comprised five core academics, and six researchers. When the need for additional experts to analyse data and user requirements was identified,
two more researchers joined the team. Team members underwent necessary security clearance before gaining access to the data and the LEA personnel. Ethics committee approvals were sought at various 
stages to govern the activities conducted throughout this project, including a qualitative user study and the analysis of the associated data.



\section{Methodology}
\label{section:methodology}

\subsection{Qualitative data collection}


We conducted qualitative research, specifically semi-structured interviews~\cite{William2015}, \update{with investigators from the LEA}.
\update{A senior manager at the LEA (the project contact) identified} three teams responsible for investigating the 
crime type for which the system is being developed \update{and invited their members to participate}. 
\update{In the end, twelve participants were recruited from the participant pool.}
Interviewing different teams (where each team worked on a distinct task) enabled us to capture diverse perspectives and expectations for the system, 
and helped us elicit a wide range of user requirements.


Prior to the interview, the participants provided informed consent, understood that they were not obliged to answer any questions they preferred not to,
and were free to withdraw at any time. At the beginning of each interview, the participants were told about the purpose of the 
research, and that their participation would be treated confidentially and anonymously. The rest of the semi-structured 
questions were divided into two parts. The first part focused on existing practices within the LEA,
while the second part focused on the participants’ needs in a system to be developed.

Semi-structured interviews \update{with open-ended questions} enabled the participants to have more control over the specific topics covered 
and allowed us the flexibility of asking follow-up questions when needed. 
For confidentiality, interviews were arranged via the project contact and the participants 
were known to the research team by their first name only. Interviews were predominantly 
conducted in person or via a video-conference, \update{with each session lasting up to 1.5 hours}.
An additional participant chose to submit an anonymous written response.  
The interviews were not recorded; instead, only notes were permitted to be taken during the interviews to capture a summary of participants’ responses (i.e. not verbatim).
\update{At least two interviewers were present, one to ask questions and the other to write down notes. More comprehensive notes were created shortly after each interview and
the findings were later discussed with the LEA for validation purposes.} No direct quotes were available for use at any time.

%
%
%
%

\subsection{Analysis of interview notes}

Notes taken by the research team during the interviews were subsequently analysed using \update{inductive} Thematic Analysis~\cite{Braun2006UsingThematicAnalysis}
by two authors of the paper, a postdoctoral researcher in software engineering
with experience in both qualitative research and the development of AI-assisted systems, and a postdoctoral researcher in 
forensic psychology with extensive experience on the project and a strong understanding of the domain. 
This integration of technical and psychological expertise was valuable for conducting the research 
and translating user needs into technical requirements.

Each researcher independently extracted requirements from the notes and any disagreements were discussed. Where necessary,
another author of the paper (with \update{extensive} expertise in requirements engineering \update{and process automation in law enforcement}) 
acted as a moderator to discuss and resolve any disagreements. 

While the researchers spent
considerable time analysing the notes from the first three interviews, discussing the terminology, identifying synonyms
and deciding on the format of the extracted requirements, the researchers became more confident and aligned with each other
for the remaining interviews. Therefore, the calculated Fleiss' Kappa \cite{fleiss1971measuring} agreement scores indicate 'Moderate' agreements
for the first three interviews and `Substantial'/`Almost perfect' agreements for the remaining interviews 
(see Table~\ref{table:interraterInterviews}). Note that only the 11\textsuperscript{th} interview included two participants
(all the others included only one) and that 
notes were taken and analysed for both participants together.

\vspace{-2mm}

\begin{table}[htp]
    \small
    \caption{Inter-rater reliability statistics across interviews.}
    \vspace{-4mm}
    \begin{center}
    \begin{tabular}{c c c c}
    \hline
    Interview notes	& Participants & Kappa	 & Interpretation\\
    \hline
    N1	& P1 & 0.42 &Moderate\\         
    N2	& P2 & 0.52 &Moderate\\         
    N3	& P3 & 0.44 &Moderate\\         
    N4	& P4 & 0.73 &Substantial\\      
    N5	& P5 & 0.74 &Substantial\\      
    N6	& P6 & 0.84 &Almost perfect\\   
    N7	& P7 & 0.71 &Substantial\\      
    N8	& P8 & 0.67 &Substantial\\      
    N9	& P9 & 0.62 &Substantial\\      
    N10	& P10 & 0.86 &Almost perfect\\  
    N11	& P11, P12 & 0.70 &Substantial\\
    \hline
    \end{tabular}
    \end{center}
    \label{table:interraterInterviews}
    \end{table}%

\vspace{-2mm}

When analysing the collected data, the researchers identified some requirements as out of scope. These requirements did not align with the agreed scope of the project
and could not be considered due to the limited time available for developing and integrating the AI-assisted system with 
predefined functionalities into the LEA's pipelines. As a result, the out-of-scope requirements were discarded.
    


After removing out-of-scope and duplicated requirements, the total of 827 unique requirements were grouped by themes 
and sub-themes. \update{For instance, some participants stated they would like to receive a summary report that includes explanations of the generated output. 
In this case, the user requirements extracted from the relevant notes would fall under the theme `tool\textquotesingle s output' and the sub-theme `explainability'.} 
Other examples of requirements can be found at \url{https://anonymous.4open.science/r/userStudy_data/}.

Grouping requirements thematically helped us have a clear understanding of data collected during the 
interviews and perform a straightforward requirement analysis. While functional themes describe functionalities that 
the AI-assisted tool should provide, the non-functional themes mostly describe quality attributes needed for 
the system to be adopted in the LEA and leveraged by its users. 

Due to the confidential nature of the research project, we are not able to discuss the functional requirements. However, the non-functional requirements
(also referred to as `Specific quality requirements' in \cite{Glinz2007On})
and their themes help us identify the key aspects and considerations the present study sought to explore.
These insights can serve as a guide for other researchers and developers working on AI-assisted decision-making tools \update{for detection and prevention of various crime 
types in this domain}.

%

\section{Findings}
\label{section:findings}

In this section, we answer our research questions by leveraging the key findings from our qualitative research conducted
in the LEA. 

\subsection{RQ1: \update{What are the key user needs driving the design of an AI-assisted decision- making system in law enforcement?} }

In law enforcement, investigators employ numerous tools to help them process criminal investigation data, such as offense data, offender data and forensic data.
However, a major challenge  in existing software is accounting for the dynamic and evolving nature of crime. 
Offenders' activities typically change over time and space~\cite{UKgovLetter2023}. 
Norms in society also change, making it crucial to build automated software that will remain relevant and worthwhile. 
Changing norms, together with the increasing volume of incoming digital data, can leave a burden on investigators to
adapt and contextualise tool outcomes, increasing the manual effort needed. 

The key findings for our RQ1 (see Table~\ref{table:RQ1_findings}) show that LEA investigators deal with \textbf{large volumes of incoming data} in their everyday processes.
A participant described the amount of data as `\textit{enormous}' and `\textit{unlimited}' (N3). As a consequence, the investigators' work is \textbf{time-consuming}
(`\textit{Working at pace is the main shortcoming currently}' (N8)), as well as 
\textbf{resource-consuming} (`\textit{Need to use a lot of resources and technology team to help}' (N11)). 

Participants in the study recognised and commented on \textbf{a need for reducing manual work} --
`\textit{If the tool could limit the amount of [data] to [process manually], that would be great'} (N3) -- especially
when handling specific crime types where frequent exposure to challenging material might have a negative psychological impact on investigators 
\cite{duran2022impact, Duran2023Associations, Woodhams2024Model}. 
Based on the note `\textit{The tool could help with the psychological impact of the [data]}' (N3),
our study confirms that there is \textbf{a need for reducing the psychological impact on investigators by reducing exposure to distressful material}.   

Due to the evolving nature of criminal activities and behaviour, some participants noted that they `\textit{constantly need to evolve [their] technology or strategies}' (N3), 
as they identified \textbf{a need for adaptation and evolution of current practices}, so 
that they can keep pace with this evolution and combat crime effectively.

It has been recognised that numerous tools are available, for example, to help investigators in various stages of data analysis
and decision-making. 
However, \textbf{existing tools have limitations}, and as the participants explained, the current tools `\textit{do not completely fit with the landscape}' (N10), 
and generate `\textit{a lot of false positives}' (N8). Finally, from N8 we learned that the participant prefers manual data
analyses, as they have `\textit{not seen any automated tool currently as effective as manual interpretation}'. These limitations of current practices and tools in use 
clearly show a need for adopting other technologies that would fit better within the LEA.

\vspace{-1mm}
\begin{tcolorbox}[colframe=black, width=0.48\textwidth, left=1mm, right=1mm, top=1mm, bottom=1mm, boxrule=0.5pt]
    The participants of our study reported many challenges in their decision-making activities, including the high volume of incoming data and the manual 
    data processing. These activities are time- and resource-intense and have a negative psychological impact on the study participants.
    The limitations of the current tools in use indicate a need for a tool that would overcome these limitations, 
    analyse a high volume of data efficiently and help the investigators in 
    their everyday processes within the LEA. 
\end{tcolorbox}
\vspace{-2mm}

\begin{table*}[!t]
\small
\caption{Summary of key findings for RQ1: \update{What are the key user needs driving the design of an AI-assisted decision-making system in law enforcement?}}
\vspace{-3mm}
\label{table:RQ1_findings}
\centering
\begin{tabularx}{\textwidth}{>{\hsize=0.38\hsize\arraybackslash}X|>{\hsize=0.62\hsize\arraybackslash}X}

\toprule
    \textbf{Key findings} & \textbf{Relevant remarks from notes taken in interviews.}\\
    \midrule
    There are high volumes of incoming data. & \makecell[cl]{`The volume of cases is enormous.' (N3) \vspace{1mm}\\
    `Unlimited volume of [data].' (N3) \vspace{1mm}\\
    `There are large datasets that are constant in daily [...] work stream.' (N8) } \\
    \hline
    \vspace{1mm}Working on data is time-consuming. \vspace{1mm}&  \vspace{1mm} `Working at pace is the main shortcoming currently.' (N8)\vspace{1mm}\\
    \hline
    Working on data is resource-consuming. & \vspace{1mm} \makecell[cl]{'Need to use a lot of resources [...] to help [make decisions].' (N3)\\
    `Need to use a lot of resources and technology team to help.' (N11)} \vspace{1mm} \\
    \hline
    \vspace{1mm}There is a need for reducing manual work. \vspace{1mm}& \vspace{1mm}`If the tool could limit the amount of [data] to [process manually], that would be great.' (N3) \vspace{1mm}\\ 
    \hline
    There is a need for reducing the psychological impact on investigators. & \vspace{1mm} `The tool could help with the psychological impact of the [data].' (N3) \vspace{1mm}\\
    \hline
    There is a need for adaptation and evolution of current practices. & \vspace{1mm} \makecell[cl]{`Each [crime] category continually evolves.' (N10) \vspace{1mm}\\
    `Technology moves so quick, sometimes one tactic [for detecting crime] works once, it might \\ not work again, so we constantly need to
    evolve our technology or strategies.' (N3)} \\
    \hline
    Existing tools have limitations. & \vspace{1mm} \makecell[cl]{
    `Current [...] tools do not completely fit with the landscape.' (N10) \vspace{1mm}\\
    `This is a tool which has been used previously to mixed effect. It does [identify]
    correctly, but \\ there are a lot of false positives.' (N8) \\
    `Not seen any automated tool currently as effective as manual
    interpretation.' (N8)}\\
    \bottomrule
\end{tabularx}
\end{table*}

\subsection{RQ2: What are the \update{primary} quality requirements for an AI-assisted decision- making system in law enforcement?}

Quality requirements play an important role in designing and developing AI-assisted systems as they directly influence how the system functions, while 
satisfying users' needs. Some of the requirements identified as important in the studies by Habibullah et al.~\cite{Habibullah2021}
(such as scalability, accuracy, justification, trust and adaptability) and Herrewijnen et al.~\cite{Herrewijnen2024Requirements} (such as understandability and explainability) were also discussed by our participants, 
while taking into account the dynamics of the domain they work in (see Table~\ref{table:RQ2_findings}).

In the law enforcement domain, when new data arrives daily, the system should be able to process large amounts of data and produce 
the output speedily. Sometimes, having a decision is urgently needed to continue with the investigation. Therefore, it is understandable
that the participants of this study showed concerns about 
the \textbf{scalability} of the system and its speed to generate the output. The notes indicate
that the speed is `\textit{a big performance concern}' (N1). 

Despite speed being a concern, the accuracy of the results has a higher priority and should be satisfied first. As we find out from N5, 
it is preferable if the system `\textit{took a longer time but had better accuracy}'.
Apart from assigning a high priority to this quality requirement, the same participant
defined \textbf{accuracy} as `\textit{the ability to understand context correctly [...] like a human can}'.

To improve the understandability and explainability of the generated output, the tool needs to justify its predictions. \textbf{Justification}
can be ensured by providing a summary report that would explain the tool's output in more detail. Additionally, a confidence score
could be provided alongside a summary report to represent the likelihood that the AI-generated output is correct.
Lately, the intelligence assessment community has been using a `probability yardstick' \cite{ProbabilityYardstick2023}
as the preferable language when talking about the likelihood of predicted events. The confidence score (traditionally presented as a numerical value from 0 to 1
or lately as a descriptive `probability yardstick' expression from `Remote chance' to `Almost certain') together with the summary report would overall strengthen the tool's \textbf{trustworthiness}.
Participants also believed that they can build trust in the tool by using it over time (a participant `\textit{would have to have used the tool
himself to have faith in it}' (N1)).

For building trust, both low and high confidence scores are important. A high score indicates that the user could trust the output, while a low score
signals that a high-confidence output could not be generated. As stated in N1, `\textit{if [the tool] is unsure [P1] would rather it says it was 
unsure}'. In this case, investigators could use tool's partial outputs, sometimes even analysing additional sources of information,
to make their decisions.

Many participants in our study described the domain of their work as very dynamic. Therefore, another required characteristic
of an AI-assisted system is \textbf{adaptability}. To keep pace with the changes
in government guidance, and the nature of criminal activities and behaviour \cite{UKgovLetter2023} that impact input data, 
an AI-assisted system needs to constantly adapt to these changes. Ideally, the system should be able to recognise and interpret
new information and predict new classes not given
at the time of training. As we find out from N8, the adaptation should happen in particular when the information not
seen before becomes common in the input data -- `\textit{You would want [the tool] to adapt, not week by week, but say if 
new [information in the input data] emerge then you would want it to consider those}'. 

In order to help the AI-system learn from the new information in the data, a participant would like to 
`\textit{interact with [the] system and drag things around}' and if needed, `\textit{the tool could learn from that interaction}' (N1).
This user-friendly interaction would enable investigators to feed the tool with the domain knowledge it can leverage when making further predictions.

\vspace{-1mm}
\begin{tcolorbox}[colframe=black, width=0.48\textwidth, left=1mm, right=1mm, top=1mm, bottom=1mm, boxrule=0.5pt]

   Our findings highlight that the dynamics of the law enforcement domain   
   give rise to key quality requirements for an 
   AI-assisted decision-making system.
    Scalability, accuracy, justification of the generated output and decisions, trust and adaptability were identified
    by the participants in our study that need to be considered in the dynamic domain they work in.
    Our study also shows that an AI-assisted tool needs human 
    involvement to be able to adapt to the changes in government guidance and in the nature of criminal activities and
    behaviour.
\end{tcolorbox}
\vspace{-2mm}

\begin{table*}[!t]
    \small
    \caption{Summary of key findings for RQ2: What are the \update{primary} quality requirements for an AI-assisted decision-making system in law enforcement?}
    \vspace{-3mm}
    \label{table:RQ2_findings}
    \centering
    \begin{tabularx}{\textwidth}{>{\hsize=0.38\hsize\arraybackslash}X|>{\hsize=0.62\hsize\arraybackslash}X}

    \toprule
        \textbf{Key findings} & \textbf{Relevant remarks from notes taken in interviews.}\\
        \midrule
        Speed is a priority as long as it does not affect the accuracy. & \makecell[cl]{
        `Speed is a big [performance concern].' (N1) \vspace{1mm} \\
        `Speed is tricky. It would be realistic for [the tool] to take some time.
        [P5 would] rather it took \\ a longer time but had better accuracy.' (N5)}\\
        \hline
        Accuracy means the ability to understand the context of the input data. &  \vspace{1mm} \makecell[cl] {
        `Accuracy would be very important, as in the ability to understand context correctly.' (N5) \vspace{1mm}\\
        `Accuracy is understanding context like a human can.' (N5) \vspace{1mm}\\
        `Context is everything. If you can build a tool that can understand the
         context [of the input \\ data], that would be useful.' (N1)}\vspace{1mm}\\
        \hline
        The tool should provide a summary report to justify output. &  
        \vspace{1mm} \makecell[cl] {'[P1] would like a summary report for all the tool's decisions.' (N1) \vspace{1mm}\\
        `The tool output should summarise why it thinks what it thinks and should provide enough \\ context for the [investigator] to fully understand
        the decision.' (N5) \vspace{1mm}\\
        `Summary report [is] sufficient: this is what was found, here are a few key
        elements you can \\ go back and review.' (N8)}\vspace{1mm}\\
        \hline
        Improving justification also strengthens trustworthiness. & \vspace{1mm} \makecell[cl]{
        `Justification for all decisions would help build trust.' (N1) \vspace{1mm}\\
        `A confidence score might give them more trust in the tool alongside a
         summary report.' (N1) \vspace{1mm}\\
        `In terms of what would make people trust the output, it would be useful for
        them to be able \\ to see why each score has been applied to each [decision].' (N9)}\\
        \hline
        The tool needs to adapt to the dynamics of the domain. & \vspace{1mm} \makecell[cl]{'We constantly need
        to evolve our technology or strategies.' (N3) \vspace{1mm}\\
        `If the government turned around and said we aren't interested in
        [particular activities] \\ anymore,
        then the tool should account for this.' (N5) \vspace{1mm}\\
        `[The tool] needs to be adaptable as we learn more.' (N4) \vspace{1mm}\\
        `You would want [the tool] to adapt, but say if new [data] emerge then you
         would want it to \\ consider [the new information in the data].'
        (N8)} \\
        \hline
        \vspace{1mm} Human-machine interaction \update{is needed for collecting users' feedback that the tool can learn from}. & \makecell[cl]{
        `[P1 would like to] interact with [the] system and drag things around.
        If the tool could learn \\ from that interaction that would be good.' (N1)}\\
        \bottomrule
    \end{tabularx}
    \end{table*}

\vspace{-2mm}

\subsection{RQ3: \update{What are the responsibilities humans need to undertake in an AI-assisted decision- making system in law enforcement?}}

Humans play central roles in investigative activities within the law enforcement domain \cite{IALEA2012}. This was confirmed in our study, as one of the participants
stated, `\textit{professional judgment plays a big role in how we do things}' (N3). 
The integration of AI-based tools to support investigative purposes thus requires a clear definition of the roles a human agent is expected to perform when using such a tool. 
The roles humans may adopt have been explored in studies aimed at developing human-in-the-loop AI solutions, and they may involve humans  
labelling data~\cite{YAN_2020_generation}, validating and correcting AI-generated decisions~\cite{Chen2021Interactive} or teaching an AI agent how to perform a task~\cite{Haug2018Haug}.

In our focus on law enforcement, we learned from the participants that `\textit{new [data] is coming in all the time}' (N9) and
 `\textit{[information in the input data] change all the time}' (N6). Therefore, it is challenging 
to have full understanding of the data all the time. Furthermore, a participant recognised
a specific type of information, as well as the context in which the information is given,
as potentially difficult for AI to understand and 
process properly (`\textit{But a human would need to assess [the input data] as
 [this type of information in the input data] may 
be lost by AI, [...] and AI might not pick this up.'} (N6)). Therefore, 
\textbf{a human is needed to assess input data that is difficult for AI to interpret} (see Table~\ref{table:RQ3_findings}).

Due to the changes in government guidance and the context of input data, features and classes used in AI predictions 
need to be reviewed regularly to ensure alignment between the AI-assisted system and the new norms and regulations.
As described by a participant, `\textit{it would be useful to have a more regular update of [classes] that
the tool can [predict].'} (N8). The same participant would also want `\textit{a wider consensus of opinion on what [classes] 
should be considered [for prediction].'} (N8).
Therefore, \textbf{regular updating of prediction classes would be useful} and
investigators are required to assess and agree on these classes
when they detect that the AI-generated predictions become irrelevant or insufficient over time.

In general, participants see AI as a good start, being able to 
recognise relevant information at a high level, but \textbf{a human needs to complete the task}, 
particularly when
it comes to taking details into account and making the final decision.
Automatically generated output should be validated by a human, as noted by some participants:
`\textit{the results that
come back would need a human to look at them and assess them}' (N9),
`\textit{A manual assessment [of the tool’s output] would need to be
applied}' (N8), and '\textit{there needs to be human resources put in place to actively
review [the tool’s output]}' (N10). Therefore,
\textbf{investigators are required to actively review and validate the AI-generated output.}

Another participant would like to `\textit{be able to manually make [a decision]}' (N1), which emphasises the
importance of keeping the option to \textbf{make decisions \update{manually}} and regardless of the automatically generated predictions. 

\textbf{Regular overall monitoring and validation of the tool is required}, as we learned from the notes 
`\textit{An annual review would be best, if possible}' (N3) and to `\textit{have someone keep checking if it is doing what it needs to do [...] maybe
every month or year}' (N6). Monitoring and validating is also seen as a chance for the tool to improve
(`\textit{If [the tool] could improve over time that would be great}' (N3)). This suggests that the tool should not only maintain strong performance, 
but also improve by integrating new knowledge from input data and domain experts.






\begin{table*}[!t]
    \small
    \caption{Summary of key findings for RQ3: \update{What are the responsibilities humans need to undertake in an AI-assisted decision-making system in law enforcement?}}
    \vspace{-3mm}
    \label{table:RQ3_findings}
    \centering
    \begin{tabularx}{\textwidth}{>{\hsize=0.38\hsize\arraybackslash}X|>{\hsize=0.62\hsize\arraybackslash}X}

    \toprule
        \textbf{Key findings} & \textbf{Relevant remarks from notes taken in interviews. }\\
        \midrule

        A human is needed to assess input data that is difficult for AI to interpret. & \vspace{1mm} `A human would need to assess [input data] as [some information in the input data] may 
be lost by AI, [...] and AI might not pick this up.' (N6) \vspace{1mm}\\
        \hline
        \vspace{1mm}Regular updating of prediction classes would be useful.\vspace{1mm} &  \vspace{1mm}
        `It would be useful to have a more regular update of [classes] that the tool can [predict].' (N8) \vspace{1mm}\\
        \hline
        AI is a good start, but a human needs to complete the task. & 
        \vspace{1mm} \makecell[cl]{`The AI is a good start, but you would want a human to take over.' (N5) \vspace{1mm}\\
        `AI good for the macro to code to a point you can say this is  a [criminal activity] 
        but would \\ want an analyst to review the more  micro details.' (N7) } \vspace{1mm}\\
        \hline
        Automatically generated output should be validated by a human. & 
        \vspace{1mm} \makecell[cl]{
        `Manual [data] checks could be automated, though the results that come back would need \\ a human to look at them and assess them.' (N9)\vspace{1mm}\\
        `A manual assessment [of the tool's output] would need to be applied.' (N8) \vspace{1mm}\\
        'There needs to be human resources put in place to actively review [the tool's output].' (N10)\vspace{1mm}\\
        `It would have to have some level of human interaction to do checks at some point – \\ \hspace{0mm} [P6] couldn't 
        hand over [the tool output] without being certain.' (N6)} \\
        \hline
        An investigator should be able to make decisions \update{manually}. & \vspace{1mm}
        `[It] would be useful to be able to manually make [a decision].' (N1) \vspace{1mm}\\

        \hline
        The tool requires regular monitoring and overall validation. & 
        \vspace{2mm} \makecell[cl]{`[P6] is happy for [the tool] to run continuously and 
        then just have someone keep checking \\ if it is doing what it needs to do [...] maybe every
         month or year.' (N6) \vspace{1mm}\\
        `If [the tool] could improve over time that would be great. An annual
         review would be best, \\ if possible.' (N3)} \\
        \bottomrule
    \end{tabularx}
    \end{table*}

\vspace{-1mm}
    
\begin{tcolorbox}[colframe=black, width=0.48\textwidth, left=1mm, right=1mm, top=1mm, bottom=1mm, boxrule=0.5pt]
    Our study identifies several responsibilities humans need to undertake in an AI-assisted system within the LEA and highlights their importance. Investigators with their domain expertise are needed for assessing the information and its context in
    the input data that might be challenging for the AI to interpret, regular updating and extending features and classes for AI predictions,
    and validating the automatically generated output. On the other hand, data scientists are needed for regular system monitoring and overall validation.
\end{tcolorbox}
\vspace{-2mm}






\section{The AI-assisted system design}
\label{section:prototype}



The \update{design} of our AI-assisted system is guided by the \update{user} requirements extracted from the
interviews with the investigators (i.e., \update{domain experts who are} end users of the system), while understanding both the limitations of the existing tools and the capabilities of 
modern technologies, \update{as well as addressing ethical considerations~\cite{Osasona2024Reviewing,Ethics2023,NPCC2023}}. Our key findings showed that the AI-assisted decision-making system should be accurate, justifiable, trustworthy and adaptable,
as well as allow interaction with users. This interaction enables the human involvement, as regular manual \update{input and output validation, 
and overall performance monitoring} are preferable in decision-making in the LEA, rather than full \update{automation}.

\begin{figure}[t]
    \caption{The human-in-the-loop AI-assisted decision-making system \update{that achieves user requirements for scalability, accuracy, justification,
    trustworthiness and adaptability.}}
    \vspace{3mm}
    \centering
    \includegraphics[width=0.9\linewidth]{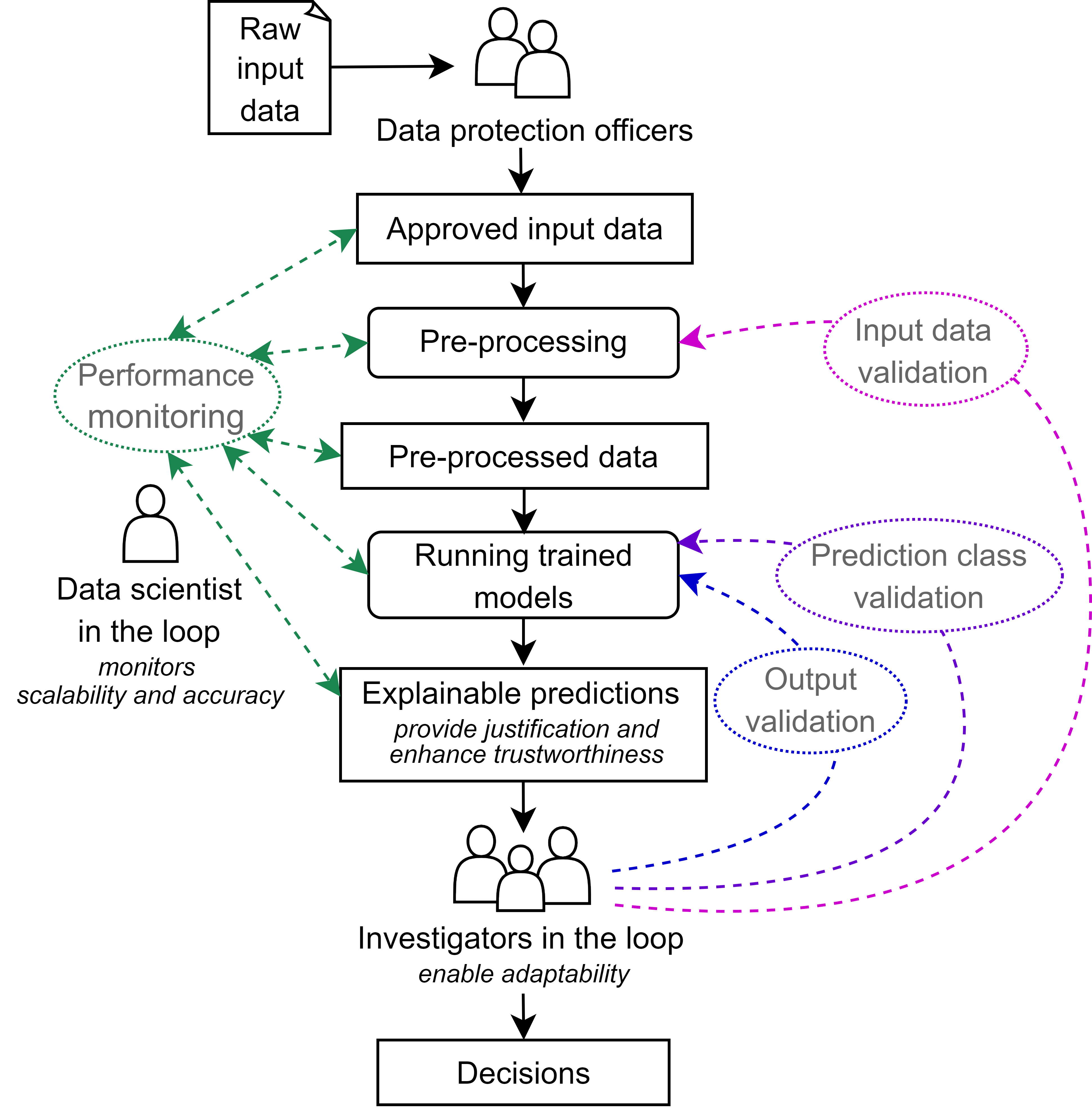}
    \label{fig:diagram}
\end{figure}

\vspace{2mm}
\update{\textbf{Human-in-the-loop AI-assisted system.}}
We adopt a standard workflow of a human-in-the-loop AI-based system~\cite{Wu2022Survey} and extend it with additional \update{responsibilities} assigned to 
\update{investigators and a data scientist}
(see Figure~\ref{fig:diagram}). \update{The input data (for both training and the deployment) needs to be approved by data protection officers, 
ensuring data privacy and the lawful data usage in the system.} \update{The approved data is then cleaned from the noise and transformed into the format
that could be fed into trained models.}

\update{Prior to model training}, the interdisciplinary academic team worked together with LEA investigators to define the set of features and \update{classes} 
of interest and manually labelled datasets with descriptions of various criminal behaviour and activities. 
Potentially suitable AI algorithms were then chosen for development, applied to the dataset to meet the requirements, and tested early.
Factors that impacted this choice included the type of data and data distribution for
classes of interest. Once the trained models reached desirable performance (\update{measured by widely-used metrics for AI-based crime prediction systems~\cite{Dakalbab2022AI}}), 
\update{the models were integrated into the prototype}.



\update{To address participants' concerns about scalability, the system generates predictions as new data becomes available, 
running in the background while end users engage in other activities (such as analysing previous batches of input data), instead of operating on user demand. 
To address accuracy, various models
were trained for the same prediction class to ensure that the positive prediction (that flags a criminal behaviour or activity) would not be missed.}

\update{Alongside the predictions from the trained models and }
in order to enhance the understandability and trustworthiness, end users get an explainable output that justifies system's predictions.
\update{The explanation helps end users validate predictions, which is crucial for the accuracy of the system.}

The system also allows end users to validate the input data that AI struggles to understand, and update features and prediction classes in the case of changes in the data or \update{government guidance}.
These human-in-the-loop \update{responsibilities} are assigned to the LEA investigators, to complement the automated process with their domain knowledge and expertise
\update{and enable the system to adapt to the changes}.
Additionally, the system provides support for regular monitoring and evaluation in the deployment. For instance, a data scientist
may regularly check performance metrics, particularly accuracy and scalability, to ensure they remain above a pre-defined threshold and update system components if needed.




\vspace{2mm}
\noindent \textbf{Addressing ethical considerations.}
The use of AI in decision-making raises various ethical concerns, such as transparency, accountability, data privacy, bias and unfairness~\cite{Osasona2024Reviewing},
which must be carefully addressed, particularly in law enforcement. 
We address transparency by providing explanations alongside the generated predictions. Regarding accountability and responsibility, 
investigators are the end users who validate the system's output
and, guided by the validated predictions, are the ultimate decision makers in our system. 
The data privacy issue is handled by data protection officers as they
need to approve both datasets used for training and the raw data used in the deployment, ensuring data privacy and the lawful data usage in the whole decision-making system.

\update{While manually analysing and labeling datasets for training models, no examples of potential bias or unfairness were identified.
However, as they are recognised as critical ethical concerns,
we address them in our human-in-the-loop design. Investigators in the loop can identify and correct any cases of biased or unfair output. 
Later, when retraining models using the investigators' feedback, the models will improve their understanding of data and reduce the possibility of biased or unfair predictions.
Additionally, we ensure the robustness of our system by training models on high-quality data from various datasets of criminal behaviour and activities, manually labelled by our interdisciplinary academic team.  
Our approach of addressing these ethical considerations aligns with the UK government guidance on leveraging ethical AI in the public sector~\cite{Ethics2023} 
and the principles established in `Covenant for Using Artificial Intelligence (AI) in Policing'~\cite{NPCC2023}.}

Note that the details about the data, the trained models and the decision-making component in our setting cannot be made public,
owing to confidentiality agreements. However, we believe that even at a high level, our \update{proposed design} still provides 
sufficient insight into the human-in-the-loop AI-based system \update{and can serve as a guide for other researchers and developers working on AI-assisted decision-making systems in this domain}.




\section{Discussion}
\label{section:discussion}


\sloppy
In this section, we discuss additional findings from our study and provide final guidance for designing and adopting AI-assisted decision-making systems in law enforcement.






\vspace{2mm}
\noindent \textbf{Accuracy alone is not enough.} 
Accuracy has been recognised as one of the most important quality requirements of AI-assisted systems ~\cite{Habibullah2021}, 
which was also confirmed by our study.
Some participants clearly highlighted the importance of accuracy
 (`\textit{accuracy would be very important}' (N5), 
 `\textit{[the software is a success] if over time it is considered by practitioners to be an accurate predictor}' (N4)).
However, when participants were asked about the most important quality attributes, notes N4 and N10 revealed that ease of use was key.
This finding aligns with our results on the human-in-the-loop roles in RQ3. Multiple human-in-the-loop roles are
required to ensure that the correctness of the system does not impact the final decision in crime detection, particularly in cases 
of a low confidence score that indicates uncertainty in the tool's output.


To address the 'easy-to-use' preference, an important focus of the development of an AI-assisted system should be on \textbf{user-friendly visualisation} of the generated output, 
while implementing \textbf{main functionalities}, satisfying \textbf{quality requirements} and taking into account \textbf{ethical considerations}.
All of this will help the AI-assisted system be widely adopted in law enforcement.


\vspace{2mm}
\noindent \textbf{The importance of human engagement.} Some researchers (e.g., Raaijmakers~\cite{Raaijmakers2019AI}) have identified the importance of human involvement 
in handling AI-generated outputs in law enforcement.
Our study aligns with this, 
as our participants argue that `\textit{a manual assessment [of the tool's output]}' (N8) is needed, and it
`\textit{would need a human to look at [the results] and assess them}' (N9).


However, in some cases, 
end users would not accept the tool's output and would have to make corrections. Instead of simply changing or discarding the output,
a participant of our study `\textit{would be happy to take the time to provide an explanation as to why [they] have changed the output}' (N8). Similarly,
another participant `\textit{is happy to spend time providing feedback to help improve the tool's future use}' (N1). This shows that the participants were
willing not only to use this tool in their everyday activities, but also to help the tool learn new patterns in data
and improve its predictions. 
Actively providing feedback on the tool's output will help the tool re-learn and adapt to changes.
Only through human engagement, an AI-assisted decision-making system will be able to \textbf{learn patterns beyond its original training and improve its predictions}.

\vspace{2mm}
\noindent \textbf{Humans will remain a key part.}
It is highly unlikely that an AI-assisted system in a LEA will  be fully automated in the foreseeable future.
The idea of fully automated AI-assisted systems triggers many discussions in the literature. 
For instance, the study conducted by Winter et al.~\cite{Winter2022Towards}
highlighted that some developers are ready to accept that automatically generated output does not always need human validation, as 
long as the tool has already proven its success, the tool's output satisfies some of the required criteria and does not have
a significant impact on other software or developers. 

However, the participants in our study seem convinced that even with the further development of AI and 
the ability of the tool to keep up with that trend, it is highly unlikely that an AI-assisted system in law enforcement will ever 
be fully automated due to the dynamic and complex nature of the domain.
Nonetheless, our study indicated that users accept that `\textit{[performing some tasks automatically] would be useful}' (N10) and
that `\textit{there are times a tool would be helpful, there are other times they would only know through human intuition'} (N5).
This shows that even with AI-based task automation, \textbf{investigators' knowledge and expertise will remain not only necessary, but also crucial in decision-making}.

\section{Threats to validity}
\label{section:threats}


\textbf{Internal validity.} 
\update{Two members of the research team with a background in forensic psychology conducted interviews with LEA participants.
To avoid any potential bias due to the interviewers' experience and background, the main and follow-up questions were discussed and prepared by the whole team
prior to the interviews.}

\update{
To address interpretive bias in the thematic analysis of interview notes, 
two researchers independently conducted thematic analyses and when necessary discussed with an academic, experienced in 
requirements engineering and process automation in law enforcement, to resolve any disagreements}.

A threat to internal validity might also arise from the ambiguity of certain parts of the interview notes presented in this paper. 
Due to restrictions around confidentiality, some notes had to be redacted, with certain information removed or presented more generally. 
However, we aimed to preserve 
the original meaning as much as possible and still use the notes as indicators of our key findings.

\vspace{1mm}

\noindent \textbf{External validity.} Our qualitative research was performed across three LEA teams \update{investigating the crime type that the AI system is designed for, but each team focused on a distinct task}.
\update{Interviewing different teams enabled us to collect valuable and various requirements to guide the design of our system.}
However, our insights may not
be applicable to investigators from other teams or other LEAs \update{and our findings do not generalise to settings different from ours}. \update{On the other hand,} we believe that this study gives valuable insights into
\update{our} AI-assisted decision-making system \update{designed} for the LEA and that 
other studies can be conducted to \update{extend our research and }investigate \update{the design of AI systems} in other settings.


\section{Related Work}
\label{section:related}

In this section, we provide a summary of related research.
\vspace{2mm}

\noindent
\textbf{User studies on AI-assisted tools in practice.}
Various studies have been conducted with the aim to understand and analyse users' views on integrating AI-based systems in their
workflows. AI seems a good fit for automated program repair and fault localisation where developers are happy to 
accept automatically-generated code changes, bug fixes or fault locations \cite{Winter2022Towards,Williams2024UserCentric, Kang2024AQuantitative}. 
Our findings overlap with some of the findings from these studies, such as that humans need to be involved and validate automatically generated outputs.
However, while AI-based program repair shows potential for full automation when additional validation techniques are employed,
decision-making in crime detection can never be fully automated due to the nature of the
domain, which requires expert knowledge for making final decisions. 


Robe and Kuttal~\cite{Robe2022} investigate how conversational AI agents
increase programming efficiency in the context of interactive pair-programming.
Our study confirms the importance of having a human alongside
an AI tool when performing complex tasks.
Furthermore, Carmona et al.~\cite{Carmona2024} report on users' expectations towards interactive learning solutions in detection of users' physical activities.
Similarly to this work, our study shows participants' willingness to provide feedback to improve the tool's learning process.

\update{Steyvers and Kumar~\cite{SteyversKumar2024Three} present challenges of designing AI-assisted decision-making systems, focusing on the domains of health, finance and autonomous driving. 
They highlight the importance of addressing challenges related to human interaction with AI systems and providing explanations for AI-generated outputs. 
This aligns with the findings of our study, showing the common requirements for AI systems across different domains.}

\vspace{2mm}

\noindent
\textbf{AI-assisted decision-making in law enforcement.}
While the studies from above show the importance of human factors and human involvement in other domains, the human-in-the-loop design of an AI system is even more desirable in 
law enforcement due to the nature of this domain.
Researchers have considered various human-in-the-loop methodologies to facilitate the adoption of AI in government and law enforcement.

Sarzaeim et al.~\cite{Sarzaeim2023} present a pipeline for  government agencies to adopt human-in-the-loop 
when automating their processes. 
Agudo et al.~\cite{Agudo2024} conducted user studies, in the field of justice, to better understand the impact of the time at which AI system 
support is presented to a human on the accuracy of the final decision.

A study by Dempsey et al.~\cite{Dempsey2023Exploring} indicates mixed views
of police officers towards AI and identifies features such as fairness, accountability, transparency and explainability to be desirable for an AI system. 
On the other hand, Herrewijnen et al.~\cite{Herrewijnen2024Requirements} present
a study where police officers expressed their positive attitudes towards AI. The aim of that exploratory study was to gain
insights into attitudes towards AI technologies in general and explainable AI more specifically.
Our work differs from this related work as we conducted a study based on user requirements 
with the aim to investigate the limitations of traditional technologies used by our participants and identify a need for AI technologies to be leveraged, as well as
defining specific roles humans need to have to support AI effectively. Our findings also indicate that end users are willing 
to help the AI system learn from their domain knowledge.
Finally, we complement the studies from above by identifying additional 
essential quality requirements for an AI-assisted system, such as scalability, accuracy, justification, trustworthiness and adaptability.



\vspace{2mm}

\noindent
\textbf{Quality attributes of AI-based systems.}
The importance of understanding and defining quality attributes for AI-assisted systems was recognised by 
Heyn et al.~\cite{Heyn2021Requirement}. The authors highlight that human factors are crucial for a system's adoption
and success and pose critical questions such as ``Will humans accept decisions of the automated system?'' and
``Will they react accordingly?'. We address these questions in our study.

Other research has further explored quality attributes for AI systems.
So \cite{So2020} proposes a human-in-the-loop design cycle framework to replace qualitative user testing. 
The main principle in that framework is that generated user feedback should serve as input for model training. 
Honeycutt et al.~\cite{Honeycutt2020} describe user experiments conducted to study the effect of interacting with an AI system that utilizes 
their feedback  on their  trust  in it, their perception of how the model changed, and how they perceived its accuracy. Their study
found that  participants  perceived the system to be less accurate and had less trust compared to those who did not 
provide feedback, regardless of the actual change in accuracy.
\update{Trustworthiness was also a focus in the study conducted by Baldassarre et al.~\cite{Baldassarre2024TrustworthyAI}, 
where they recommended the steps to follow when developing trustworthy AI applications.}

Van der Stappen and Funk~\cite{Stappen2021}  present a study in which participants interacted with a variety of interfaces and assessed them using interaction quality and efficiency metrics. 
 Based on this, the authors proposed a set of design guidelines and provide a unified means for evaluating  interaction experience and system learning efficiency. 
Habibullah and Horkoff~\cite{Habibullah2021} conducted a series of interviews with software developers to understand their perceptions on non-functional requirements
in a machine learning context. Our research on the other hand explores users' perceptions on crucial requirements
for AI-based decision-making systems within a specific domain.



\section{Conclusion}
\label{section:conclusion}

In this paper, we presented the key findings of qualitative research conducted within 
a LEA. The participants of our study highlighted that they work with large volumes of digital data 
daily and that there is a need for a tool that uses modern technologies to efficiently analyse data
and make predictions. They also argued that the tool needs to satisfy essential quality requirements, 
such as scalability, accuracy, justification, trustworthiness and adaptability, to be accepted by end users.
Additionally, they see the AI-assisted tool as an adjustable and evolvable system that fits the dynamics 
of their domain, particularly changes in the input data and rules for making decisions.
Finally, being user-friendly will help the tool to be adopted in the LEA.

The AI-assisted system in the LEA requires humans being involved in the process of 
decision-making, in particular to validate the input data and tool's output and revisit prediction classes,
as well as to monitor and validate the performance of the overall system.
The human engagement is seen as essential and enduring, as it is highly unlikely that the tool will ever achieve full automation.

The findings of our user study and our experience in developing a prototype of a human-in-the-loop AI-assisted system
can guide other developers when building an AI-assisted system in the same domain, or a domain with similar dynamics and complexity.



\begin{acks}
We are very grateful to the LEA for supporting our research and
the participants of our study for sharing their time, expertise and insights with us.
This work has been partially supported by CHEDDAR: Communications Hub for Empowering Distributed ClouD Computing Applications and Research
funded by the UK EPSRC under grant numbers EP/Y037421/1 and EP/X040518/1.
\end{acks}

\newpage

\bibliographystyle{IEEEtran}
\bibliography{bib}

\end{document}